\documentclass{aa}

\input psfig.sty

\begin{document}
\newcommand{\mo}{$M_{\odot}$}
\newcommand{\mob}{$M_{\odot}\;$}

\title{Habitable Zones of Host Stars During the Post-MS Phase}

\author{Jianpo Guo\inst{1,2,3}
        \and Fenghui Zhang\inst{1,2}
        \and Zhanwen Han\inst{1,2}
        }

\offprints{Guo et al.}

\institute{National Astronomical Observatories/Yunnan Observatory,
Chinese Academy of Sciences,
           Kunming, 650011, P.R. China\\
           \email{guojianpo1982@hotmail.com}
          \and
          Key Laboratory for the Structure and Evolution of
Celestial Objects, Chinese Academy of Sciences,
           Kunming, 650011, P.R. China\\
           \and
           Graduate School of the Chinese Academy of Sciences,
Beijing, 100049, P.R. China\\
          }

\date{Received 16 Nov. 2009; accepted 9 Feb. 2010}

\abstract{A star will become brighter and brighter with stellar
evolution, and the distance of its habitable zone will become
farther and farther. Some planets outside the habitable zone of a
host star during the main sequence phase may enter the habitable
zone of the host star during other evolutionary phases. A
terrestrial planet within the habitable zone of its host star is
generally thought to be suited to life existence. Furthermore, a
rocky moon around a giant planet may be also suited to life survive,
provided that the planet-moon system is within the habitable zone of
its host star. Using Eggleton's code and the boundary flux of
habitable zone, we calculate the habitable zone of our Solar after
the main sequence phase. It is found that Mars' orbit and Jupiter's
orbit will enter the habitable zone of Solar during the subgiant
branch phase and the red giant branch phase, respectively. And the
orbit of Saturn will enter the habitable zone of Solar during the
He-burning phase for about 137 million years. Life is unlikely at
any time on Saturn, as it is a giant gaseous planet. However, Titan,
the rocky moon of Saturn, may be suitable for biological evolution
and become another Earth during that time. For low-mass stars, there
are similar habitable zones during the He-burning phase as our
Solar, because there are similar core masses and luminosities for
these stars during that phase.

 }

\titlerunning{HZs of Stars during Post-MS}

\authorrunning{Guo et al.}

\maketitle

\section{Introduction}
Typically, stellar habitable zone (HZ) is defined as a region near
the host star where water at the surface of a terrestrial planet is
in the liquid phase (e.g., Kasting et al. 1993; Franck et al. 2000;
Noble et al. 2002; Jones et al. 2006). The inner HZ boundary is
determined by the loss of water via photolysis and hydrogen escape.
And the outer HZ boundary is determined by the condensation of
carbon dioxide crystals out of the atmosphere (von Bloh et al.
2007).

Previously, it was generally paid attention to the HZs of host stars
during the main sequence (MS) phase, because the evolution from
biochemical compounds to primary life needs long enough time. And a
terrestrial planet around a host star covers a presumed heavy
bombardment phase as on Earth (Jones 2004; Lal 2008), during the
initial evolutionary stage of the host star. And the heavy
bombardments make the temperature on the surface of a terrestrial
planet become very high, which is not suitable for biological
evolution. Only when the temperature on the surface of a terrestrial
planet goes down, the evolution from biochemical compounds to
primary life may start. These mean that the evolutionary age of a
host star when life comes into being on a terrestrial planet around
the host star is longer than the evolutionary timescale from
biochemical compounds to primary life on the terrestrial planet.

The evolutionary timescale of a host star during the MS phase is
generally much longer than the timescale of heavy bombardments on a
planet around the host star. Therefore, all the planets around a
host star had become stable, when the MS phase of the host star is
terminated. Without heavy bombardments, the temperature on the
surface of a planet is mainly decided by the luminosity of its host
star and the distance from the planet to the host star. Generally
speaking, a host star becomes brighter and brighter during the post
main sequence (post-MS) phase, mainly including subgiant branch
phase, red giant branch (RGB) phase, horizontal branch (HB) phase
and asymptotic giant branch (AGB) phase. And some terrestrial
planets outside the HZ of a host star during the MS phase will enter
the HZ of the host star during the post-MS phase. Once these
terrestrial planets enter the HZ of the host star, the evolution
from biochemical compounds to primary life may start immediately.
The timescale of the evolution from biochemical compounds to
prokaryote life in habitable environment may be as short as about
100 million years (Watson 2008). And all the evolutionary timescales
of low-mass stars (defined in subsection 3.2) during the subgiant
branch phase, the RGB phase and the HB phase are longer than 100
million years. Furthermore, positional changes of the HZ with
stellar age are considered in future studies of long-term changes of
the planetary biospheric conditions (Franck et al. 2000; Noble et
al. 2002). Hence, it is meaningful to study the HZs of host stars
during the post-MS phase.

Life-seeds may also migrate from one terrestrial planet to another
(Buccino et al. 2007). A 1.9 kg meteorite (ALH84001) discovered in
1984 in Allan Hills, Antarctica, is believed to have been blasted
off the Martian surface due to an asteroid or comet impact about 15
million years ago (McKay et al. 1996; Hoyle and Wickramasinghe
1999a). There are also some meteorites found on Earth, whose source
is the Moon (Warren 1994). These proves that a major meteorite
impact on a planet or a moon would eject fragments into space and
the fragments may finally fall down another planet. If there are
life-seeds (such as spore, fungi, bacterial and plant seed) in the
fragments, they can be ejected into space. Some microbes may also
survive after a long journey through space from one planet to
another planet (Joseph 2009; Wickramasinghe et al. 2009). And some
plant seeds resist deleterious conditions found in space, e.g.,
ultra low vacuum, extreme temperatures and intense ultraviolet
light. In a receptive environment, life-seeds could liberate a
viable embryo, viable higher cells or a viable free-living organism
(Tepfer and Leach 2006). Therefore, life can transfer from one
planet to another through ejected fragment. In addition, life seeds
may also be saved on comets and can be sent to remote places by the
comets, and the life seeds may take root, provided that the
conditions they visit on become habitable (Hoyle and Wickramasinghe
1999a, 1999b, 1999c; Wickramasinghe et al. 2009).

Generally speaking, a terrestrial planet in the HZ of a host star is
suited to life existence. Moreover, a rocky moon orbiting a giant
planet or a brown dwarf could also be habitable, provided that the
planet-moon system or the brown dwarf-moon system is within the HZ
of the host star (Williams et al. 1997; Lal 2010). There are several
rocky moons in solar system, such as Europa, Ganymede and Titan.
These rocky moons will be habitable when Solar becomes brighter
enough during the RGB phase or the HB phase. And there may be lives,
which can generate from biochemical evolution or be transferred from
other planets or moons, living on the rocky moons during that time.
Furthermore, it is even thought that there may be life existence on
these moons now, considering the potential of extremophiles to
survive in highly inhospitable environments on Earth (Lal 2008).

Using the boundary distances of HZ (Jones et al. 2006), we calculate
the HZs of host stars during the post-MS phase. It is found that
Mars' orbit, Jupiter's orbit and Saturn's orbit will enter the HZ of
Solar during the subgiant branch phase for about 1.48 Gyr, the RGB
phase for about 175 million years and the HB phase for about 137
million years, respectively. For low-mass stars, there are similar
HZs as Solar during the HB phase. This means that a plant may also
enter the HZ of a low-mass star during the HB phase, provided that
the distance from the planet to the host star is just equal to the
distance from Saturn to Solar. For intermediate-mass stars (defined
in subsection 3.3), the HZs of host stars during the He-burning
phase become farther and farther, with the stellar masses
increasing.

The outline of the paper is as follows: we describe our methods in
Section 2, show our results in Section 3, present some discussions
in Section 4, and then finally in Section 5 we give our conclusions.

\section{Methods}
\subsection{Input physics about stellar evolution}
We use the stellar evolution code of Eggleton (1971, 1972, 1973),
which has been updated with the latest input physics over the last
three decades (Han et al. 1994; Pols et al. 1995, 1998). We set the
convective overshooting parameter, $\delta_{\rm OV}=0.12$ (Pols et
al. 1997; Schr\"{o}der et al. 1997). We also take Reimers' type
mass-loss (Reimers 1975) into account, with Reimers's parameter
$\eta=1/4$. In our calculation, the value of metallicity is 0.02 and
stellar mass is from 0.80 to 4.00 \mob (Guo et al. 2009).

We adopt the metal mixture by Grevesse and Sauval (1998). We use
OPAL high temperatures opacity tables (Iglesias and Rogers 1996;
Eldridge and Tout 2004) in the range of
$4.00<\mathrm{log(}\mathit{T}\mathrm{/K)}\leq 8.70$, and the new
Wichita state low temperature molecular opacity tables (Ferguson et
al. 2005) in the range of
$3.00\leq\mathrm{log(}\mathit{T}\mathrm{/K)}\leq 4.00$. And we have
made the opacity tables match well with Eggleton's code (Chen and
Tout 2007; Guo et al. 2008).
\subsection{Boundary distances of HZ}
Jones et al. (2006) gave the flux at both the inner and the outer HZ
boundaries, as a function of effective temperature ($T_{\rm eff}$).
\begin{equation}
\frac{S_{\rm in}}{S_{\odot}}=4.190\times10^{-8}T_{\rm
eff}^{2}-2.139\times10^{-4}T_{\rm eff}+1.296, \label{mdis}
\end{equation}
\vspace{-5.0mm}
\begin{equation}
\frac{S_{\rm out}}{S_{\odot}}=6.190\times10^{-9}T_{\rm
eff}^{2}-1.319\times10^{-5}T_{\rm eff}+0.2341, \label{mdis}
\end{equation}
where $S_{\odot}$ is solar constant and $T_{\rm eff}$ is in Kelvin.

As $L=4{\pi}d^{2}S$, and the three parameters are just as
$L_{\odot}$, AU and $S_{\odot}$, for our Earth. Hence, the distances
at both the inner and the outer HZ boundaries are given by
\begin{equation}
\frac{d_{\rm in}}{\rm AU}=\bigg[\frac{L/L_{\odot}}{S_{\rm
in}/S_{\odot}}\bigg]^{1/2}, \label{mdis}
\end{equation}
\begin{equation}
\frac{d_{\rm out}}{\rm AU}=\bigg[\frac{L/L_{\odot}}{S_{\rm
out}/S_{\odot}}\bigg]^{1/2}. \label{mdis}
\end{equation}
\section{Results}
\subsection{HZ of Solar during the post-MS phase}
Let us firstly define the termination of the main sequence (TMS) for
a star, when hydrogen is exhausted in the center of the star. This
is noted as an asterisk on the Hertzsprung-Russell (H-R) diagram of
our Solar in Fig. 1. We then calculate the HZ for our Solar from TMS
to the tip of the RGB phase, as seen in Fig. 2. At the start of TMS,
the orbit of Mars is just within the HZ of our Solar and remains so
for the first about 1.48 Gyr. The temperature on the surface of Mars
is moderate and water can remain to be in the liquid phase.
\begin{figure}
\psfig{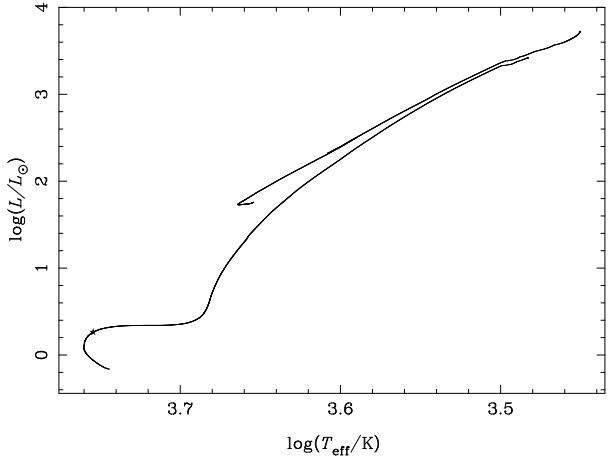}
\caption{H-R diagram of Solar.} \label{ised}
\end{figure}
\begin{figure}
\psfig{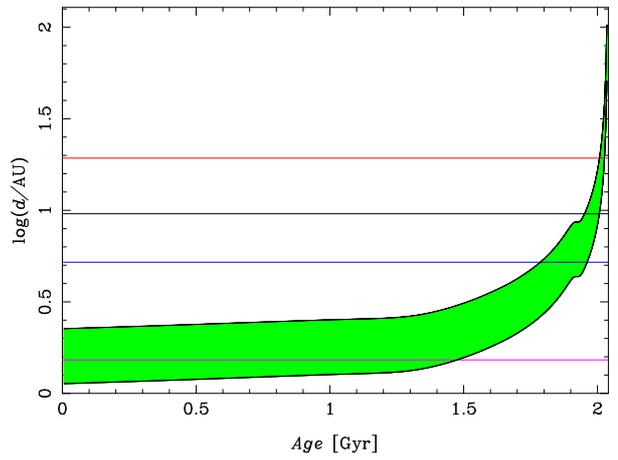}
\caption{HZ of Solar from TMS to the tip of the RGB phase. The four
horizontal lines are the semimajor axes of Mars, Jupiter, Saturn and
Uranus from bottom to top, respectively.} \label{ised}
\end{figure}

Jupiter' orbit is within the HZ of our Solar during the RGB phase
for about 175 million years. Life is unlikely at any time on
Jupiter, as it is a giant gaseous planet. However, the rocky moons
of Jupiter, such as Europa and Ganymede, may be suitable for
biological evolution and life may come forth on the rocky moons
during that time. Furthermore, life seeds may also migrate from
other planets with lives to Europa or Ganymede, the two rocky moons
of Jupiter may become another two Earths during that time.

The He-burning phase for a star generally contains two phases, just
as HB phase and AGB phase. During the HB phase, there is a
He-burning core for a star and the change of the luminosity is not
evident. And the luminosity during the whole HB phase is near to the
luminosity at the zero age horizontal branch (ZAHB), for low-mass
stars. By contrast, Helium is exhausted in the center of a star
during the AGB phase, there is a He-burning envelope for the star
and the change of the luminosity is great and rapid. Generally
speaking, the evolutionary timescale of a star during the AGB phase
is much shorter than that of the star during the RGB phase. We give
the age-luminosity diagram of our Solar during the He-burning phase,
the ZAHB is noted as a point and the start of the AGB phase (just as
the termination of the HB phase) is noted as a square, as seen in
Fig. 3. It is found that the luminosity changes tinily and the
evolutionary timescale is 136.89 million years for our Solar during
the HB phase. On the contrary, the luminosity changes evidently and
the evolutionary timescale is only 9.81 million years for our Solar
during the AGB phase.
\begin{figure}
\psfig{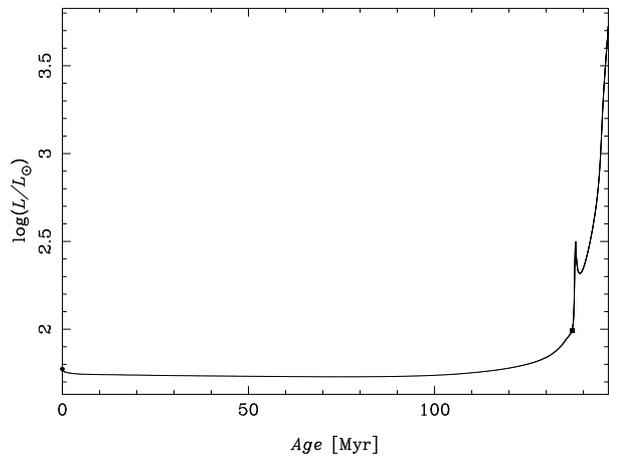}
\caption{Luminosity of Solar during the He-burning phase.}
\label{ised}
\end{figure}

As we have get the $T_{\rm eff}$, the luminosity and the
evolutionary timescale of our Solar during the He-burning phase, we
calculate the HZ of Solar during the He-burning phase, as seen in
Fig. 4. It is found that the HZ of Solar during the whole HB phase
is near to the HZ of Solar at the ZAHB, as the luminosity of Solar
changes tinily during the HB phase. By contrast, the HZ of Solar
during the AGB phase changes greatly and rapidly, as the luminosity
of Solar changes greatly and rapidly during the AGB phase. Saturn's
orbit is within the HZ of Solar during the whole HB phase, and also
within the HZ of Solar at the start of the AGB phase for 0.31
million years. Therefore, Saturn's orbit is within the HZ of Solar
during the He-burning phase for 137.20 million years. Life is
unlikely at any time on Saturn, as it is also a giant gaseous
planet. However, Titan, the rocky moon of Saturn, may be suitable
for biological evolution and become another Earth during that time.
And Uranus' orbit is within the HZ of Solar during the AGB phase for
only 5.48 million years, which is so short that the evolution from
biochemical compounds to primary life can not be accomplished.
\begin{figure}
\psfig{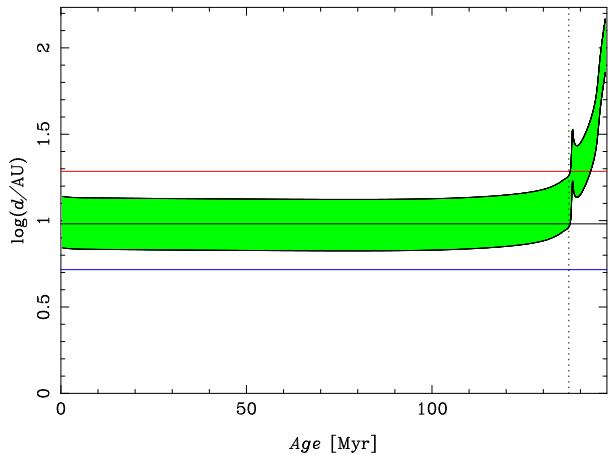}
\caption{HZ of Solar during the He-burning phase. The vertical
dotted line is the boundary between the RGB phase and the AGB phase.
And the three horizontal lines are the semimajor axes of Jupiter,
Saturn and Uranus from bottom to top, respectively.} \label{ised}
\end{figure}
\subsection{HZs of low-mass stars during the He-burning phase}
The low-mass stars in this paper are defined as the stars which
develop degenerate helium cores on the RGB phase and ignite helium
in a degenerate flash at the tip of the RGB phase (Hurley et al.
2000). The maximum masses of low-mass stars with different
metallicities are different. In our calculation, the value of
metallicity is 0.02. Hence, the maximum mass of low-mass stars is
about 2.00 \mo, which is also the minimum mass of intermediate-mass
stars. And the core masses for all low-mass stars during the
He-burning phase have similar values, just as about 0.47 \mob in our
calculation. Therefore, low-mass stars during the He-burning phase,
also including HB phase and AGB phase, have similar $T_{\rm eff}$
and luminosities, as seen in Fig. 5.
\begin{figure}
\psfig{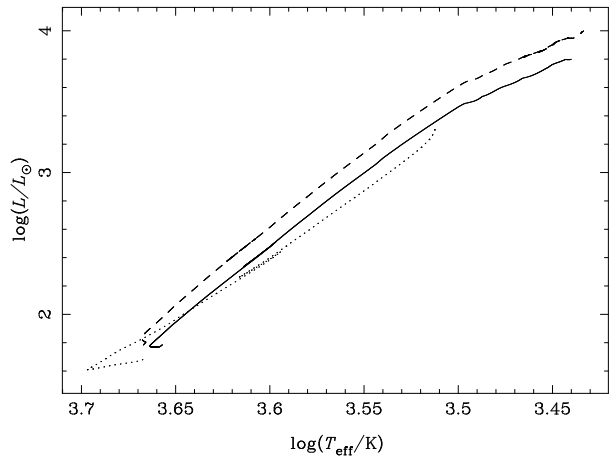}
\caption{H-R diagrams of low-mass stars during the He-burning phase,
with masses 0.80 (dotted line), 1.20 (solid line) and 1.50 \mob
(dashed line).} \label{ised}
\end{figure}

We calculated the HZs of low-mass stars during the He-burning phase,
with masses 0.80, 1.20 and 1.50 \mo, as seen in Fig. 6. There are
also similar HZs for the three low-mass stars during the He-burning
phase. As the three host stars have similar core masses, $T_{\rm
eff}$ and luminosities during the He-burning phase. It is found that
Saturn's orbit is also within the HZs for all the three host stars
during the HB phase for more than 130 million years. It means that a
plant is also in the HZs for all the low-mass stars during the HB
phase, provided that the distance from the planet to its host star
is just equal to the distance from Saturn to Solar.
\begin{figure}
\psfig{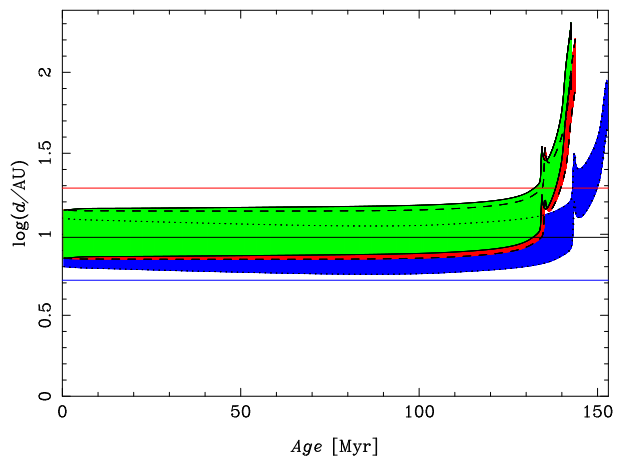}
\caption{HZs of low-mass stars during the He-burning phase, with
masses 0.80 (bottom), 1.20 (middle) and 1.50 \mob (top). The two
dotted lines, the two dashed lines and the two solid lines are the
inner and the outer boundaries of the HZs around the host stars with
masses 0.80, 1.20 and 1.50 \mo, respectively. And the three
horizontal lines have the same meanings as in Fig. 4.} \label{ised}
\end{figure}
\subsection{HZs of intermediate-mass stars during the He-burning phase}
The intermediate-mass stars in this paper are defined as the stars
which evolve to the RGB phase without developing degenerate helium
cores, also igniting helium at the tip of the RGB phase (Hurley et
al. 2000). The core masses of intermediate-mass stars during the
He-burning phase become bigger and bigger, with the stellar masses
increasing. Therefore, the luminosities of intermediate-mass stars
during the He-burning phase become higher and higher, and the
evolutionary timescale become shorter and shorter, with the stellar
masses increasing.

We calculated the HZs of intermediate-mass stars during the
He-burning phase, with masses 2.50 and 3.00 \mo, as seen in Fig. 7.
The evolutionary timescales for the stars with masses 2.50 and 3.00
\mob during the HB phase are about 169 and 91 million years,
respectively. And Saturn's orbit is within the HZ of the host star
with mass 2.50 \mob during the HB phase. However, the distance from
Solar to Saturn is shorter than the distance from the host star with
mass 3.00 \mob to the HZ of the host star during the HB phase. It
means that the boundary distances of the HZs for intermediate-mass
stars during the HB phase become farther and farther, with the
stellar masses increasing.
\begin{figure}
\psfig{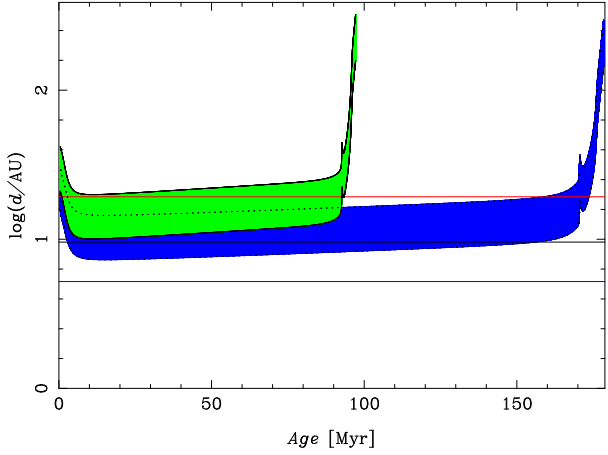}
\caption{HZs of intermediate-mass stars during the He-burning phase,
with masses 2.50 (bottom) and 3.00 \mob (upper). The two dotted
lines and the two solid lines are the inner and the outer boundaries
of the HZs around the host stars with masses 2.50 and 3.00 \mo,
respectively. And the three horizontal lines have the same meanings
as in Fig. 4.} \label{ised}
\end{figure}
\section{Discussion}
During the RGB phase, low-mass stars will remarkably lose masses
(Catelan 2009). And stellar mass loss will also cause the planet's
orbit spirally outward by conservation of angular momentum (Tarter
et al. 2007) and the law of universal gravitation. Therefore, it
meets $M_{\rm ZAMS}R_{\rm ZAMS}=M_{\rm ZAHB}R_{\rm ZAHB}$; where
$M_{\rm ZAMS}$ and $R_{\rm ZAMS}$ are stellar mass and the semimajor
axis of a planet's orbit for the host star at the ZAMS, $M_{\rm
ZAHB}$ and $R_{\rm ZAHB}$ are stellar mass and the semimajor axis of
the planet's orbit for the same host star at the ZAHB. For our solar
system, $M_{\rm ZAMS}=1.00 M_{\odot}$, $R_{\rm ZAMS}=9.576\rm AU$
(semimajor axis of Saturn's orbit) and $M_{\rm ZAHB}=0.8778
M_{\odot}$, so that $R_{\rm ZAHB}=10.909\rm AU$. And the distance of
HZ for Solar at the ZAHB is from 6.966 to 13.825 AU. Hence, Saturn's
orbit is also within the HZ of Solar at the ZAHB, taking the outward
movement of planet's orbit into consideration.

The strong ultraviolet radiation can induce DNA destruction and
cause damage to most of the biological systems (Buccino et al.
2006). And the $T_{\rm eff}$ of low-mass stars during the post-MS
phase is not high. Therefore, the UV radiation is not strong in the
HZs of low-mass stars during the post-MS phase (Guo et al. 2010),
which is suitable for biological evolution on terrestrial planets or
rocky moons in the HZs.

There is no terrestrial planet in the HZ for our Solar during the HB
phase. However, the situation may be different for other low-mass
stars. Using the data of Ida and Lin (2005) and the correlative
method of Guo et al. (2009), we calculate the probability of
terrestrial planets within the HZs around different mass stars at
the ZAHB. For the host star with mass 1.00 \mo, the probability is
0.203; for the host star with mass 1.50 \mo, the probability is
0.346. These mean that there may be lives living on terrestrial
planets around some low-mass stars during the HB phase.
\section{Conclusion}
Firstly, we calculate the HZ of Solar during the post-MS phase. It
is found that Mars' orbit, Jupiter's orbit and Saturn's orbit will
enter the HZ of Solar during the subgiant branch phase, the RGB
phase and the HB phase, respectively. These mean that Titan will be
suited to biological evolution, when Solar evolves to the HB phase.
Secondly, we calculate the HZs of low-mass stars during the
He-burning phase. As there are similar core masses, $T_{\rm eff}$
and luminosities for low-mass stars during the HB phase, there are
also similar HZs for them. Therefore, Saturn's orbit is also within
the HZs for all low-mass stars during the HB phase. Thirdly, we
calculate the HZs of intermediate-mass stars during the He-burning
phase. It is found that the HZs of host stars become farther and
father with the stellar masses increasing. Finally, we present
discussions about the outward movement of planet's orbit, the UV
radiation and the terrestrial planets within the HZs of low-mass
stars during the HB phase. One may also send any special request to
\it{guojianpo1982@hotmail.com} \normalfont{or}
\it{guojianpo16@163.com}\normalfont{.}

\begin{acknowledgements}
This work is supported by the National Natural Science Foundation of
China (Grant Nos. 10773026, 10821061 and 2007CB815406), the Chinese
Academy of Sciences (Grant No. KJCX2-YW-T24) and Yunnan Natural
Science Foundation (Grant No. 06GJ061001).
\end{acknowledgements}

{}
\end{document}